\documentstyle[11pt,newpasp,twoside,psfig]{article}
\index{anomalous X-ray pulsars}
\index{anomalous X-ray pulsars!1E~1048.1--5937}
\index{anomalous X-ray pulsars!1E~1841--045}
\index{anomalous X-ray pulsars!1E~2259+586}
\index{anomalous X-ray pulsars!4U~0142+61}
\index{anomalous X-ray pulsars!associations with supernova remnants}
\index{anomalous X-ray pulsars!AX~J1845--0258}
\index{anomalous X-ray pulsars!magnetic fields}
\index{anomalous X-ray pulsars!RXS~J170849--400910}
\index{neutron stars!magnetars}
\index{neutron stars!magnetic fields}
\index{RXTE}
\index{soft gamma-ray repeaters}
\index{soft gamma-ray repeaters!associations with supernova remnants}
\index{soft gamma-ray repeaters!SGR~0526--66}
\index{soft gamma-ray repeaters!SGR~1806--20}
\index{soft gamma-ray repeaters!SGR~1900+14}
\index{supernova remnants!Magellanic Clouds}
\index{timing!phase-coherent}
\index{XMM-Newton}
\index{Chandra}
\index{ASCA}

\def\gapp{\ifmmode\stackrel{>}{_{\sim}}\else$\stackrel{>}{_{\sim}}$\fi}
\def\lapp{\ifmmode\stackrel{<}{_{\sim}}\else$\stackrel{<}{_{\sim}}$\fi}

\def\edcomment#1{\iffalse\marginpar{\raggedright\sl#1\/}\else\relax\fi}
\reversemarginpar

\begin{document}
\title{Soft Gamma Repeaters and Anomalous X-ray Pulsars: Together Forever}
 \author{Victoria M. Kaspi}
\affil{Physics Department, McGill University, 3600 University St., Montreal, QC H3A 2T8, Canada}
\affil{Center for Space Research, MIT, 70 Vassar St., Cambridge, MA 02139, USA}

\begin{abstract}
I review of the observational properties of Soft Gamma Repeaters (SGRs)
and Anomalous X-ray Pulsars (AXPs), two unusual
manifestations of neutron stars.  I summarize the reasoning for SGRs being ``magnetars,''
neutron stars powered by the decay of a very large magnetic field,
and the now compelling evidence that SGRs and AXPs are in fact
members of the same source class, as predicted uniquely by the magnetar model.  I discuss 
some open issues in the magnetar model, and the prospects for future work. 
\end{abstract}

\section{Introduction}

Since Baade \& Zwicky made their now-famous 1934 prediction regarding the
existence of neutron stars, these amazing objects have not ceased to
surprise us in the variety of their observational manifestation.  Apart from thermal
X-ray emission from initial cooling, predicted early on and now detected in
a small handful of sources, the emission properties of neutron stars have been
formally unpredicted, and informally unimagined.

The objects today being identified as ``magnetars'' are no exception.
These sources literally exploded onto the astronomy scene in March 5,
1979, when the object today known as SGR~0525$-$66 emitted a
soft-gamma-ray burst so intense that it saturated every gamma-ray
detector that saw it (Mazets et al. 1979), likely measurably affected
the Earth's ionosphere, and implied an awe-inspiring $>10^6$ Eddington
luminosities.  This and the other handful of known ``Soft Gamma
Repeaters'' (SGRs) prompted model explanations that ranged from
vibrating neutron stars to strange star/pulsar phase transitions.
Duncan \& Thompson (1992), and quasi-simultaneously, Paczy\'nski (1992),
came up with the magnetar hypothesis, summarized below, which,
particularly following seminal papers by Thompson \& Duncan
(1995, 1996), has uniquely stood the tests of increasingly constraining
SGR observations.  They also identified ``Anomalous X-ray
Pulsars'' (AXPs) as additional members of the magnetar club.  Though at
the time having little in obviously common with SGRs, 
the AXPs, as we discuss below, have recently revealed
themselves to be true siblings of the SGRs, with so many properties in
common that the question to be answered today is ``what differentiates them
from SGRs?'' 

\section{The Observational Properties of Soft Gamma Repeaters}

\begin{table}[t]
\caption{Summary of Known and Candidate SGRs and AXPs}
\begin{tabular}{lccccc}\tableline
Name      & $P$   & $\dot{P}^a$ & $\tau_c^b$ & $B^c$ & SNR? \\
          & (s)   &  ($\times 10^{-11}$) &  (kyr)   & ($\times 10^{14}$ G) &   \\\tableline
\multicolumn{6}{c}{SGRs$^d$} \\\tableline
SGR 0525$-$66 & 8     &  ?  &   ?  &  ?  & N49 \\
SGR 1627$-$41 & 6.4?     &  ?  &   ?  &  ?  & no \\
SGR 1801$-$23 & ?     &  ?  &   ?  &  ?  & ? \\
SGR 1806$-$20 & 7.5   & 2.8 & 4.2 & 4.6 & no \\
SGR 1808$-$20? & ?    &  ?  &   ?  &  ?  & ? \\
SGR 1900+14   & 5.2   & 6.1 & 1.3 & 5.7 & no \\\tableline
\multicolumn{6}{c}{AXPs$^e$} \\\tableline
CXOU J0110043.1$-$721134? & 8.0 & ? & ? & ? & no \\
4U 0142+62 & 8.7 & 0.20 & 69 & 1.3 & no \\
1E 1048.1$-$5937 & 6.4 & 3.3 & 3.0 & 4.7 & no \\
RX 1708$-$4009 & 11.0 & 1.9 & 9.2 & 4.6 & no \\
XTE J1810$-$197? & 5.5 & 1.1? & 7.6? & 2.5? & no \\
1E 1841$-$0450 & 11.8 & 4.1 & 4.6 & 7.0 & Kes 73 \\
AX 1845$-$0258? &  7.0 &  ? & ?  & ?  & G29.6+0.1 \\
1E 2259+586 & 7.0 & 0.048 & 230 & 0.6 & CTB 109 \\ \tableline \tableline
\end{tabular}
Note that ``?'' denotes an unknown or uncertain entry.\\
$^a$Long-term average value. 
$^b$Characteristic age estimated from $P/2\dot{P}$. \\
$^c$Surface dipolar magnetic field estimated from $3.2\times 10^{19} (P\dot{P})^{1/2}$~G. \\
$^d$References for all SGRs can be found in Hurley (2000), except for the candidate SGR 1808$-$20, which was reported by Lamb et al. (2003).\\
$^e$References for all AXPs can be found in Mereghetti et al. (2002), except for CXOU J0110043.1$-$721134
which was reported by Lamb et al. (2002), and XTE J1810$-$197, reported by
Markwardt et al. (2003).\\
\end{table}

There are currently 4 and possibly 6 SGRs known (see Table 1), all but
one of which are in the Galactic plane, the exception being in the
Large Magellanic Cloud.  The latter is in the direction of the
supernova remnant N49, yet none of the others has been conclusively
linked to a remnant (Gaensler et al. 2001).  SGRs have as their
hallmark repeating short ($\sim$100~ms) soft-gamma-ray and X-ray
bursts, which have typical energies $\sim 10^{41}$~erg, and rise times
on the order of $\sim$10~ms (e.g. G\"og\"us et al. 1999).  Burst spectra
are generally well modelled by optically thin thermal bremsstrahlung
models with $kT \simeq 20-50$~eV (e.g. G\"og\"us et al. 1999).  SGR bursting
behaviour is highly episodic, with years of inactivity and weeks in
which hundreds of bursts are detected.  Occasionally, SGRs exhibit
giant gamma-ray bursts having energies $>4\times10^{44}$~erg and luminosities
$>4 \times 10^{44}$~erg~s$^{-1}$, well beyond the expectations of any
distribution of the smaller bursts.  Only two such giant bursts have
been observed, the first in 1979 from SGR~0525$-$66 (Mazets et al. 1979), and the second
from SGR~1900+14 in 1998 August (Hurley et al. 1999).

The three SGRs which have been securely seen to pulse have periods that range from
5 to 8~s.  Two of the three exhibit these pulsations in quiescence in X-rays,
and have clearly been shown to be spinning down (Kouveliotou et al. 1998, 1999).  
Under the assumption of simple magnetic dipole braking in a vacuum, the periods and
period derivatives imply surface fields of $\sim 10^{15}$~G.
The pulsations have broad profiles.  At the time of the
1998 giant burst of SGR 1900+14, its
pulse profile abruptly changed from having multi-peaked
structure to being much simpler (Kouveliotou et al. 1999).  SGR spectra in
quiescence are much softer than are the bursts, with
power laws of photon index $2-3$ typical.  They are noisy rotators, resisting phase-coherent
timing over spans longer than a few weeks
(Woods et. al. 2002).  The bursting and rotational behavior do
not appear well correlated, although at the time of the
1998 event, SGR 1900+14 showed evidence for a possible
step down in frequency (Woods et al. 1999).  There are no confirmed
detectons of any SGRs outside the X-ray or soft-gamma-ray band.

For more detailed reviews of SGRs, see Hurley (2000) or Thompson (2001).

\section{SGRs as Magnetars}

Thompson \& Duncan (1995) presented many lines of reasoning pointing to
the SGR bursts and quiescent emission being powered by the active decay
of an ultra-high magnetic field.  Here we summarize briefly the major
points:

\noindent
1. {\it From the 1979 event, which came from the direction of a
supernova remnant and in which the decaying light curve was modulated
by an 8-s periodicity, a high $B$ field is needed to slow down a
neutron star from a $\sim$10~ms period at birth within the $\sim
10^4$~yr lifetime of a supernova remnant.} This argument predicted the
subsequently observed spin downs of SGRs 1806$-$20 and 1900+14
which has provided the magnetar model
with its greatest support.  A possible spin down of SGR 0525$-$66 has
also been reported (Kulkarni et al. 2003).

\noindent
2. {\it An energy source well beyond what is available from either
rotation or accretion is needed.} The luminosity of the bursts and
quiescent emission is orders of magnitude greater than the $\sim
10^{33}$~erg~s$^{-1}$ available from rotation.  Accretion is not viable
given the apparent absence of any companion, the fact that such
behaviour is unseen in any known accreting system, and because the
accreting plasma would have to be very ``clean'' (i.e.  pure
photon/pair) to ensure low enough scattering depth for the hard burst
spectra.  For the energy of the giant bursts to be a small fraction of
available magnetic energy, $B \gapp 10^{15}$~G is required.

\noindent
3. {\it For confinement of the emission in hyper-Eddington bursts, a
magnetar strength field is required.}  The light curves of the giant bursts
include a hard, rapid-rise initial spike, followed by a quasi-exponential
decay on a time scale minutes.  The energy in the tail
was much several times that in the initial spike, implying
an event that released energy which was somehow confined for the hundreds
of seconds.  Magnetic confinement, if $B \sim 10^{15}$~G, can do it.

\noindent
4.  {\it  $B >10^{14}$~G can reduce the Thomson cross section and allow
hyper-Eddington bursts.}  The strong field greatly suppresses the
electron cross section and the consequent decrease in scattering
opacity allows higher fluxes to escape.

\noindent
5. {\it A $\sim 10^{14} - 10^{15}$~G field can undergo active decay in
a neutron star.}  Goldreich \& Reisenegger (1992) showed that fields
with strengths below these values are not expected to decay, while
those above do via ambipolar diffusion in the core and Hall drift in
the crust on $\sim 10^4$~yr time scales.

In the magnetar picture, the quiescent X-rays originate from the surface,
a result of internal heating from the decaying magnetic field.  The bursts
result from crust cracking under magnetic stress.

\section{Anomalous X-ray Pulsars}

The nature of anomalous X-ray pulsars (AXPs) was a mystery since the
discovery of the first example (Fahlman \& Gregory 1981).  There are
currently only five confirmed AXPs, all of which are in the Galactic
plane, with two at the centers of supernova remnants.  Additionally
there are three AXP candidates under investigation.  All are listed in
Table 1.  AXPs exhibit X-ray pulsations with periods in the range
6--12~s, with luminosities in the range $\sim
10^{33}-10^{35}$~erg~s$^{-1}$.  All are observed to be spinning down,
with no evidence for Doppler shifts of the pulse periods.  They have
broad pulse profiles, and X-ray spectra that are soft compared to most
accreting X-ray pulsars' and best described by two components (usually taken to be a
blackbody having $kT \simeq 0.4$~keV plus a hard power-law tail having
photon index $2.5-4$).  Several AXPs have now been detected at
optical/IR wavelengths.  This
is discussed in detail elsewhere in the proceedings (Israel et al.).
For a lengthier review of AXPs, see Mereghetti et al. (2002).

AXPs were dubbed ``anomalous'' because it was unclear what powers their
radiation.  Rotation is insufficient by orders of magnitude in most
sources.  AXPs were long thought to be accreting from a low-mass
companion (e.g. Mereghetti \& Stella 1995).  However this model is
difficult to reconcile with observations: the absence of Doppler
shifts, the absence of a detectable optical/IR companion, the apparent
associations with supernova remnants, that AXP spectra are very
different from those of known accreting sources, and that $L_x$ is
generally smaller than in known accreting sources, all are inconsistent
with this scenario.  Chatterjee et al. (2000) and Alpar (2001)
considered a model in which AXPs are accreting from disks made of
supernova fall-back material.  In this case the similarity of source
properties with those of SGRs is coincidental as no bursting mechanism
is proposed, and in any case, recent optical/IR observations do not
favour this possibility (see Israel et al., this volume).

Thompson \& Duncan (1996) suggested that the main source of free energy
for AXP emission is from the high magnetic fields ($10^{14}-10^{15}$~G)
inferred from the rotation under standard assumptions.  The implied low
characteristic ages (Table 1) are supported by the associations with
supernova remnants, and from the location of AXPs in the Galactic
plane.  The identification of AXPs with magnetars was more recently supported by
the similarity of AXP emission to that of
SGRs in quiescence; specifically, they have similar pulse periods,
spin-down rates, and quiescent X-ray spectra.  As of 1998, the only major
distinction between the properties of SGRs and AXPs appeared to be
that SGRs exhibited bursts while AXPs did not.  However,
other small distinctions exist: on average, the AXP spectra are softer
than are those of the SGRs (e.g. Kulkarni et al. 2003); the SGRs are
noisier rotators than the AXPs (Woods et al. 2002; Gavriil \& Kaspi
2002); the frequency of association with a supernova remnant is higher
for AXPs (Gaensler et al. 2001); and SGRs on average have higher
inferred $B$ fields than those of AXPs (Table 1).

\section{SGR-Like Bursts from AXPs}

As part of a major, long-term project to monitor all confirmed AXPs using
the Proportional Counter Array aboard the {\it Rossi X-ray Timing Explorer} (see
Gavriil \& Kaspi 2002 and references therein), we have recently discovered
SGR-like bursts from the direction of two AXPs.

The first discovery was of two bursts, separated by 16 days, from the
direction of 1E 1048.1$-$5937 (Gavriil, Kaspi, \& Woods 2002).
Specifically, their fast rise times, short durations, hard spectra
relative to the quiescent emission, fluence and probably clustering,
are all SGR burst hallmarks.  Note that the origin of the bursts could
not unambiguously be proven to be the AXP, given the large PCA
field-of-view, and the absence of any other radiative or spin change in
the source.  Intriguingly, the first burst's spectrum was not well fit
by a continuum model, showing evidence for a strong emission line at
$\sim$14~keV.

\begin{figure}[t]
\centerline{\psfig{file=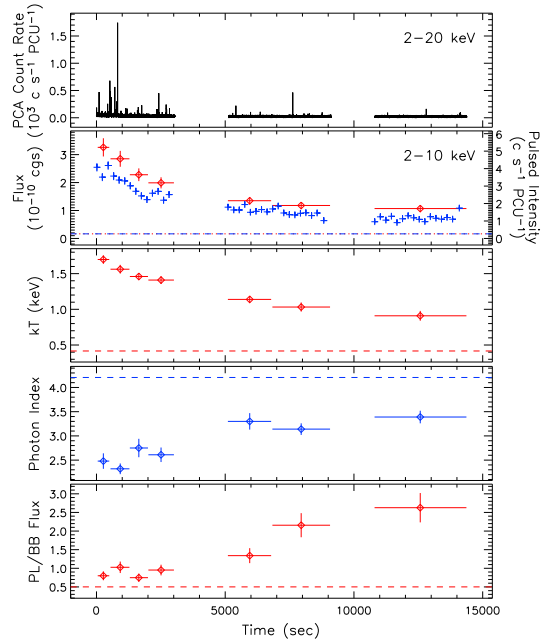,width=7cm}}
\caption{Light curve and evolution of persistent and pulsed emission
during the 1E 2259+586 outburst.  Top panel: 2--20~keV light curve at
0.125-s resolution.  2nd panel: unabsorbed persistent (diamonds) and
pulsed (crosses) fluxes (2--10~keV).  The horizontal dashed (dotted)
lines denote the quiescent levels of each parameter.  3rd panel:
blackbody temperature of the persistent and pulsed emission assuming a
two-component (blackbody and power-law) model.  4th panel:  power-law
photon index of the persistent and pulsed emission.  5th panel: ratio
of the unabsorbed 2--10~keV power-law flux and the bolometric blackbody
flux (from Kaspi et al. 2003).} 
\end{figure}

The second discovery was unambiguous:  in 2002 June, we detected over
80 bursts from 1E~2259+586 in a span of $\sim$15~ks (Kaspi et al.
2003).  Figure 1 shows the light curve, as well as several properties
of the persistent and pulsed emission during and following the
outburst.  Practically {\it every} aspect of the pulsed emission
changed:  the flux showed a large (order of magnitude) enhancement with
fast (few day) and slow (months) decay components; the spectrum
hardened but recovered within $\sim$3 days; the pulsed morphology
changed during the outburst but relaxed back to near its pre-outburst
shape after $\sim$1~week; and the pulsed fraction decreased during the
outburst to $\sim 2/3$ of its pre-outburst value, but recovered
within $\sim$6~days.  Furthermore, the pulsar suffered a possibly
resolved rotational glitch, consisting of a sudden spin up ($\Delta P/P
= 4.2 \times 10^{-6}$), followed by a large (factor of $\sim$2)
increase in the absolute magnitude of the spin-down rate (Fig. 2).  See
Woods et al. (2003) for a detailed analysis of all the above changes.
In addition, an infrared enhancement was observed immediately
post-outburst (Kaspi et al. 2003).

\begin{figure}[t]
\centerline{\psfig{file=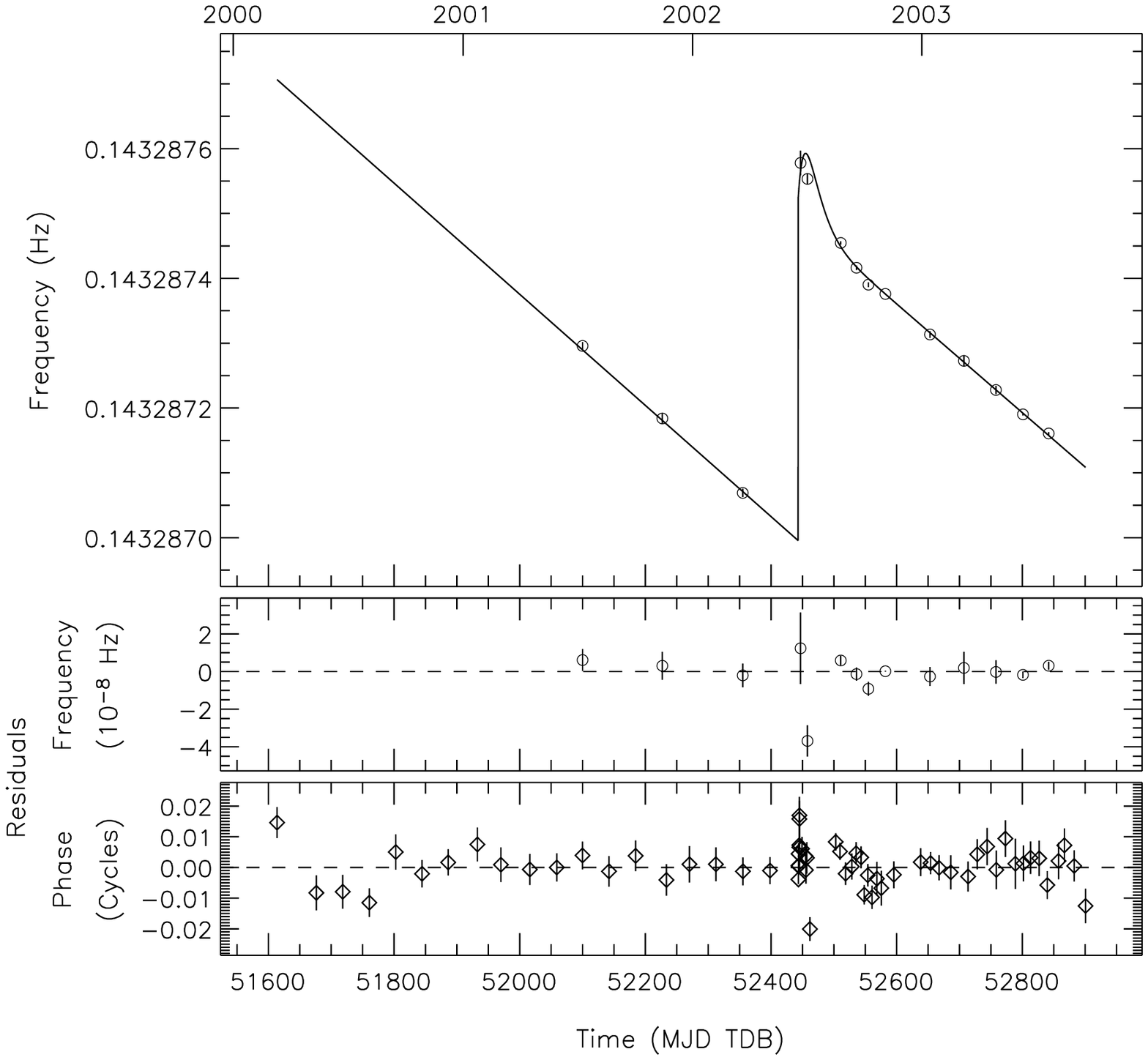,width=7cm}}
\caption{
Frequency evolution of 1E 2259+586 around the 2002 outburst for a model
including an extended exponential rise and fall in frequency
post-glitch.  Top panel: the solid line represents the best-fit
model.  The circles denote frequency measurements.  Middle panel:
the frequency residuals.
Bottom panel: phase residuals with respect to the best-fit model
(from Woods et al. 2003).
}
\end{figure}

Overall, the properties of the outburst in 1E~2259+586 argue that the
star suffered a major event that was extended in time and had two
components, one tightly localized on the surface of the star (i.e.  a
fracture or a series of fractures) and the second more broadly
distributed (possibly involving a smoother plastic change).  The glitch
points toward a disturbance within the superfluid interior while the
extended flux enhancement and pulse profile change suggest an
excitation of magnetospheric currents and crustal heating.

This AXP outburst was qualitatively and quantitatively similar to those
of SGRs.  However, there were some notable differences that may be
clues to the physical differences between the two source classes.
Specifically: the AXP bursts exhibit a wider range of durations and,
unlike SGR bursts, occur preferentially near pulse maxima; the
correlation between burst fluence and duration seen for SGRs is flatter
than for SGRs; the AXP bursts are on average less energetic than are
SGR bursts; and the more energetic AXP bursts have the hardest spectra
-- the opposite of what is seen for SGRs (Gavriil, Kaspi \& Woods, in
preparation).  Furthermore, in stark contrast to SGRs, the energy
detected in bursts ($6 \times 10^{37}$ erg, 2--60~keV) was much smaller
than that in the post-outburst persistent flux enhancement ($2\times
10^{41}$~erg, 2--10~keV).  This could indicate bursting activity that
was missed by our observations and the gamma-ray monitors (Woods et al.
2003).  No matter what, this ``quiet'' outburst strongly suggests there
are many more such objects in the Galaxy than was previously thought.

\section{Conclusions and Open Issues}

The discovery of SGR-like bursts from 1E 1048.1$-$5937 and especially 1E~2259+586
solidifies the common nature of AXPs and SGRs as predicted uniquely by
the magnetar model.  This model has now made two major predictions, namely
the spin-down of SGRs and the common nature of AXPs and SGRs, both of
which have been unambiguously borne out by observations.
The magnetar hypothesisthus appears to be very compelling.

However, there is still no direct evidence for the
magnetar strength fields.  Such evidence could in principle be obtained
from the detection of cyclotron lines in SGR or AXP spectra.  Spectral
features detected in some SGR and AXP bursts may well be providing us
with an important clue (Ibrahim et al. 2002; Gavriil et al. 2002) but
their interpretation remains unclear as of yet.

Further, one expects a magnetar/radio pulsar connection,  This could
come in two ways.  One is to detect radio pulsations from an AXP or
SGR.  Such detections have been claimed but not confirmed (see paper by
XXX, this volume).  However, detecting radio pulsations may be
impossible; the long spin periods imply small polar caps, hence
very narrow radio beams.  Alternatively, QED processes at high $B$,
such as photon splitting, may preclude the electron/positron cascades
necessary to produce radio emission (Baring \& Harding 2001).  Another
way to prove a magnetar/radio pulsar connection is to detect enhanced
X-ray emission from a high-$B$ radio pulsar.  This has not yet been
done even though several radio pulsars (see McLaughlin et al., this
volume) have now been found having inferred $B$ comparable to or
higher than that of 1E~2259+586, yet with no evidence for excess X-ray
emission.  This is puzzling, but may simply reflect that our $B$
estimate is, in reality, not very accurate.  Continued discoveries of
high-$B$ radio pulsars should prove interesting.

\acknowledgments
The author acknowledges funding from NSERC (Discovery Grant
and Steacie Fellowship), Canada Research Chairs, NATEQ, CIAR, and NASA.

\end{document}